\documentclass[10pt,twocolumn,letterpaper]{article}

\usepackage{cvpr}      % To produce the REVIEW version
\usepackage{graphicx}
\usepackage{amsmath}
\usepackage{amssymb}
\usepackage{booktabs}
\usepackage{cite}
\usepackage{epstopdf}
\usepackage{csquotes}
\usepackage{amsmath} % for math environments
\usepackage{colortbl} % for row colors
\usepackage{xcolor} % for color definitions
\usepackage{array}   
\usepackage{booktabs}
\usepackage{multirow}
\usepackage{amssymb} 

\newcommand{\rom}[1]{\romannumeral #1}
\usepackage[pagebackref,breaklinks,colorlinks]{hyperref}

\usepackage[capitalize]{cleveref}
\crefname{section}{Sec.}{Secs.}
\Crefname{section}{Section}{Sections}
\Crefname{table}{Table}{Tables}
\crefname{table}{Tab.}{Tabs.}

\begin{document}
	
	\title{MetaFE-DE: Learning Meta Feature Embedding for Depth Estimation from Monocular Endoscopic Images}
	\author{Dawei Lu, Deqiang Xiao*, Danni Ai, Jingfan Fan, Tianyu Fu, Yucong Lin\\Hong Song, Xujiong Ye, Lei Zhang, Jian Yang*\\} 
\maketitle 

%%%%%%%%% ABSTRACT  
\begin{abstract} 
	
Depth estimation from monocular endoscopic images presents significant challenges due to the complexity of endoscopic surgery, such as irregular shapes of human soft tissues, as well as variations in lighting conditions. Existing methods primarily estimate the depth information from RGB images directly, and often surffer the limited interpretability and accuracy. Given that RGB and depth images are two views of the same endoscopic surgery scene, in this paper, we introduce a novel concept referred as ``meta feature embedding (MetaFE)", in which the physical entities (e.g., tissues and surgical instruments) of endoscopic surgery are represented using the shared features that can be alternatively decoded into RGB or depth image. With this concept, we propose a two-stage self-supervised learning paradigm for the monocular endoscopic depth estimation. In the first stage, we propose a temporal representation learner using diffusion models, which are aligned with the spatial information through the cross normalization to construct the MetaFE. In the second stage, self-supervised monocular depth estimation with the brightness calibration is applied to decode the meta features into the depth image. Extensive evaluation on diverse endoscopic datasets demonstrates that our approach outperforms the state-of-the-art method in depth estimation, achieving superior accuracy and generalization. The source code will be publicly available. 

\end{abstract}

%%%%%%%%% BODY TEXT     

\begin{figure}[t]
	\centering    
	\includegraphics[width=1.0\linewidth]{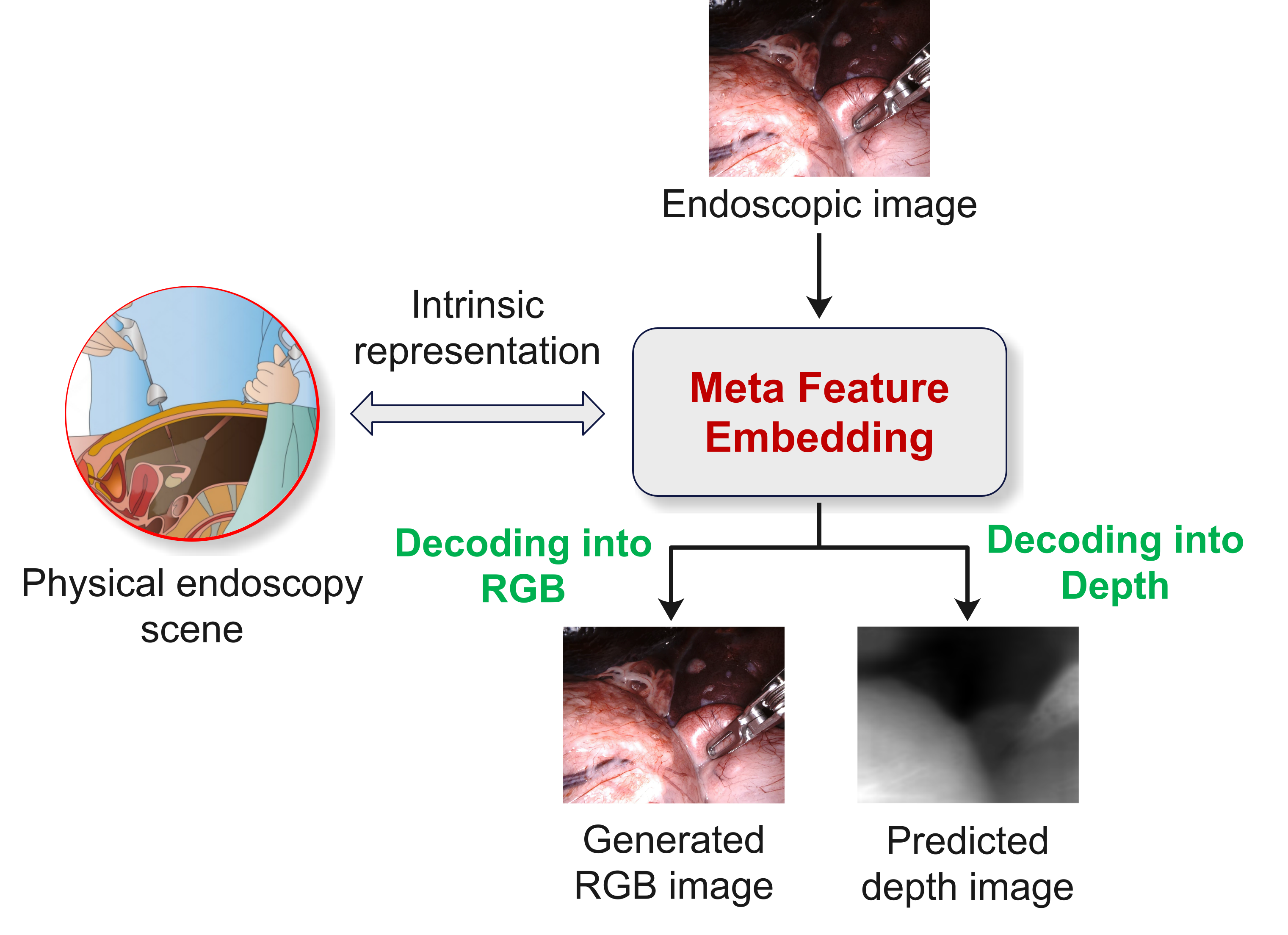}
	%\includegraphics[width=0.8\linewidth]{fig1.png}
	%\caption{A simple yet effective framework to validate the physical-world representation hypothesis by first extracting authentic features representations from RGB image generation tasks and then decoding these features into depth information. }  %
	  \caption{This paper proposes the MetaFE that represents physical entities in the endoscopic surgical scene, providing a comprehensive description of the complex surgical environment. This features can be decoded into either RGB or depth image, with the potential to generate more accurate depth estimation.} 
	\label{fig:idea_overview}  
\end{figure}   

\section{Introduction} 
\label{sec:intro}  

The 3D reconstruction of endoscopic images is a key challenge in endoscopic surgical navigation. The methods like stereo reconstruction \cite{top_scared_1}, structure from motion (SfM) \cite{sfm}, shape from shading (SfS)\cite{sfs}, and simultaneous localization and mapping (SLAM) \cite{slam} have demonstrated accurate reconstruction of sparse point clouds in target areas. However, their low computational efficiency renders them unsuitable for the time-sensitive demands of intraoperative applications. Deep learning based methods provide fast, accurate, and dense depth estimation approaches\cite{dl_1,dl_2,dl_3,dl_4,dl_5}. However, these methods primarily  address some issues in endoscopic scenes, such as lighting imbalance and sparse textures, they do not fully explore the representative features for accurate depth decoding. Nevertheless, the physical scene in endoscopic surgery is complex and cannot be fully captured by a purely modality transfer task (e.g., converting RGB images to depth images). For depth estimation, simply incorporating regularization terms in the loss function to address data-specific challenges limits further advancements in model performance. \par
Based on our preliminary experiments on endoscopic RGB image generation, we find that conditioning on depth maps in image generation tasks significantly enhances the quality of generative images (see Appendix Fig. \ref{fig:compare_depth}, Table. \ref{fig:with_depth_condition}). This observation suggests an alignment between RGB and depth image, implying that intrinsic features from enhanced generative tasks may, in turn, produce more accurate depth maps. We hypothesize, therefore, that RGB and depth images exhibit both complementarity and correlation when they capture the same endoscopic surgery physical scene from different views. In this study, we refer it as ``meta feature embedding (MetaFE)", where the representation of different modalities (e.g., RGB and depth image) derived from the same physical scene exists. Our goal is to explore this latent space and the intrinsic alignment between different visual cues, delving deeper to understand the process of decoding these features into accurate endoscopic depth images (Fig. \ref{fig:idea_overview}). \par 

\begin{figure}[t]
	\centering    
	\includegraphics[width=1.0\linewidth]{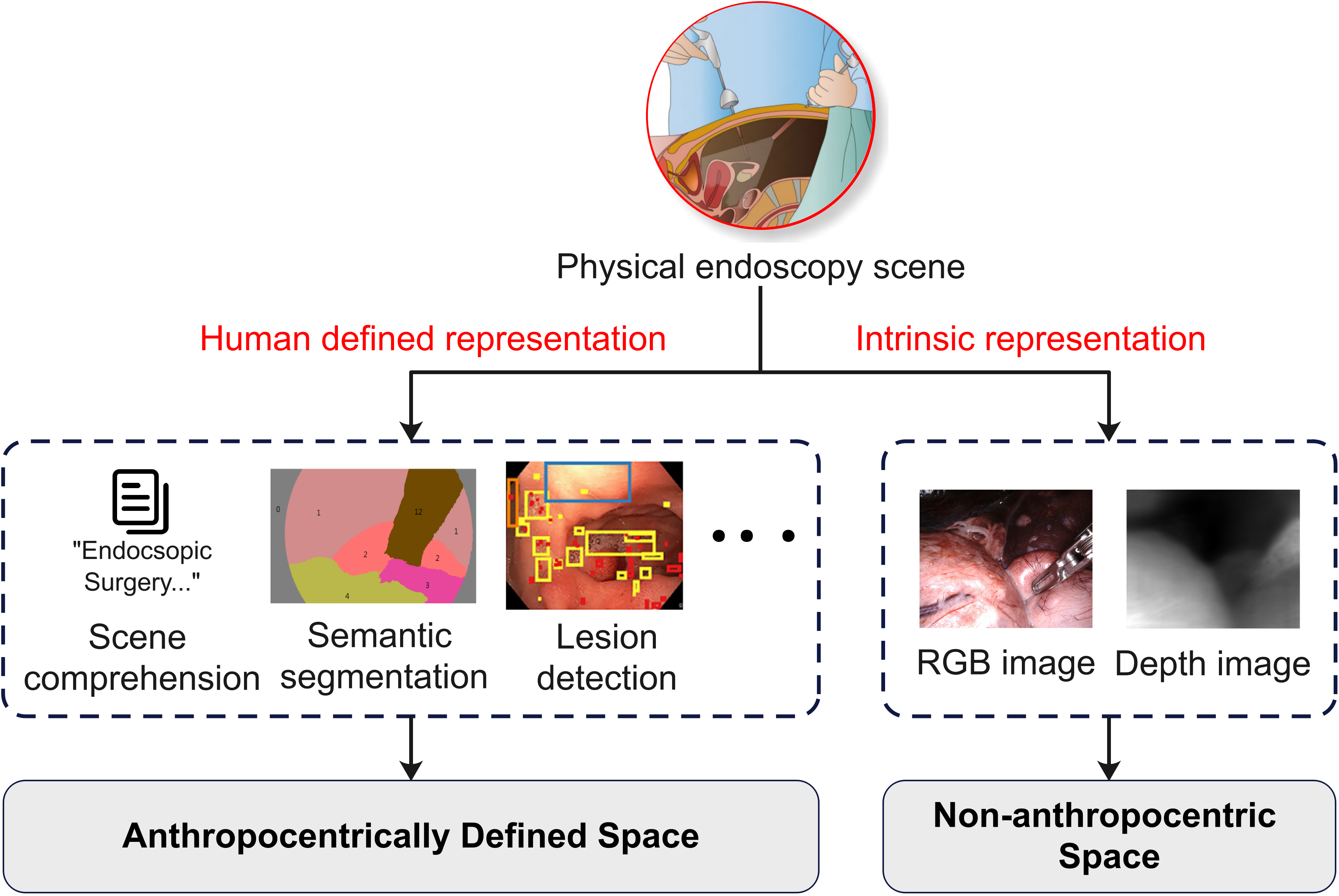}	\caption{Prior studies\cite{platonic} suggest that the text and image jointly represent the same entity in the physical world. This paper, however, categorizes modalities into non-anthropocentric space and anthropocentrically defined space based on their susceptibility to human cognition. For example, the modalities such as RGB and depth image are unaffected by human cognition, thus they are able to reflect the intrinsic physical properties.}      
	\label{fig:two_space_defination}  
\end{figure}   

To achieve this goal, we aim to answer two primary questions:  (\rom{1}) \textbf{What is the MetaFE and how can it be acquired?} More specifically, we seek to identify the latent space features that could act as MetaFE, encompassing intrinsicintrinsic properties of the endoscopic surgery physical scene, beyond modality-specific features. (\rom{2}) \textbf{How can MetaFE be applied in endoscopic image depth estimation?} More specifically, we aim to explore the method for decoding these aligned features to produce accurate depth image, enabling a more accurate interpretation of the scene. \par  

For the first question, we define the meta features as: (\rom{1}) The features are learned through self-supervised learning manner, and (\rom{2}) they are deconstructed or decoded into various visual modalities, such as RGB, depth images, surface normal and so on, for use in downstream applications. Inspired by Plato’s Allegory of the Cave, Huh et al. \cite{platonic} hypothesize a physical-world representation between RGB image and text content. As shown in Fig. \ref{fig:two_space_defination}, the reflection of physical entities in endoscopic surgery is differentiated by the presence or absence of human-defined elements. Therefore, it is reasonable to treat the feature embedding, which are simultaneously referenced by both RGB and depth images, as the space that reflects the inherent features of the physical entity itself. In this study, we employ an image generative model as a self-supervised learning pipeline to capture latent visual features and explore meta-features, due to the lack of labeled data for endoscopic monocular depth estimation. Specifically, the generative diffusion model benefits from the denoising process, which facilitates learning and enhances the induction of visual representations. More importantly, generation tasks within the same modality do not involve modality conversion and require no information from auxiliary modules (such as pose networks, lighting correction networks, etc.), nor do they require any labeled data. To take advantage of the diffusion model in alignment with previous research, we employ the latent diffusion model (LDM) with temporal information conditioned for training the generation task. 
%In this study, we designate features from diffusion model as the base of MetaFE. 

For the second question, we verify the feasibility of MetaFE by positing that features learned from raw pixels in generative tasks can be directly decoded into accurate depth images. Unlike previous work\cite{rw_dl_2,rw_dl_3}, we explore and learn the MetaFE by coupling temporal conditioning with spatial cues from frames. In practice, this coupling and fusion process is reframed as an alignment task between latent features derived from different learning pipelines. Specifically, we conduct cross normalization\cite{cross_norm} to align the distributions of the temporal diffusion and spatial features, we define the aligned features as meta features. Since both features exist in the latent space but are generated through different mechanisms, this approach aligns their distributions while preserving the maximum amount of original information. With the method reported in \cite{bright}, we utilize depth decoding with the brightness calibration to interpret the meta features into depth image.\par  

Based on the concept of MetaFE, we propose the meta feature embedding learning for depth estimation (MetaFE-DE), and the main contributions in this study are summarized as follows:     
\begin{itemize}
	
	\item We reveal the feasibility of MetaFE, where features correspond to a unique physical entity in endoscopic surgery, independent of any specific modality, and can be alternatively decoded into RGB or depth image. We demonstrate its effectiveness in endoscopic image depth estimation. 
	
	\item We provide a new learning paradigm based on the concept of MetaFE, which requires meta-features to be extracted only once, with subsequent focus solely on decoding them into the task-specific modality, without the need for task-specific features extraction.
	
	\item We conduct extensive experiments on various endoscopic image datasets, achieving new state-of-the-art performance in endoscopic image depth estimation. We reveal that different visual tasks (decoding to RGB or depth image) share common features within the abstract layers and are conducted through the same decoder pathway. 
	
\end{itemize}
%-------------------------------------------------------------------------  

\section{Related Works}  

\subsection{Diffusion Model and Representation Learning}
Diffusion models are regarded as multi-level DAEs with varied noise scales, which inherently capture meaningful representations within a latent space\cite{dif_rep_survey,ldm,INDM,EBM_34,EBM_57,EBM_68}. Therefore, leveraging the features learned during the diffusion process to effectively train downstream tasks\cite{rep_1,rep_2,rep_3,rep_4,rep_5}, such as segmentation and classification, proves both meaningful and advantageous\cite{dif_rep_cla_1,dif_rep_cla_2,dif_rep_cla_2,dif_rep_seg_2,dif_rep_multi_task,dif_rep,dif_rep}. However, while recent studies (See more in Appendix \ref{rw1}) primarily focus on methods for effectively leveraging features from the diffusion denoising process, few delve deeply into what representation learning truly entails or why it is effective.\par   

\subsection{Modality Alignment}
Huh et al. \cite{platonic} hypothesize the modalities involved in training data are shadows on the \enquote{cave wall}, which is mentioned in Plato’s Allegory of the Cave. Tian et al. \cite{real_world_1} try to align the different modalities within the contrastive loss, and believe that the more views of physical-world involved in training, the better representation captured. Zimmermann et al. \cite{real_world_2} investigate the connection between contrastive learning, generative modeling, and nonlinear independent component analysis to reveal the alignment of implicit features. Inspired by these findings (see more in Appendix \ref{model-alignment}), we aim to explore the existence of feature embeddings learned from modality alignment and underlying principles, which we refer to as the MetaFE in this study.\par 

\subsection{Endoscopic Monocular Depth Estimation}

In order to overcome the absence of depth annotation, Zhou et al. \cite{rw_dl_1} propose a self-supervised approach that reformulates depth estimation as a view synthesis problem using warping methods. This framework includes both a DepthNet and a separate PoseNet, along with a predictive mask to handle challenging scenarios like object movement and occlusion/disocclusion. This foundation has led to the development of a range of refined optimization strategies\cite{rw_dl_2,rw_dl_3,rw_dl_4,rw_dl_5,rw_dl_6,rw_dl_7}.  Considering the challenge posed by minimally invasive surgical settings, such as inconsistent interframe brightness, limit the direct applicability of these methods to endoscopic images. Shao et al. \cite{bright} propose AF-Net in order to rescue the illumination-invariant by introducing optical flow estimation module. Yang et al. \cite{rw_dl_8} introduce the LiteMono framework to enhance computational efficiency in endoscopic depth estimation. Shao et al. \cite{rw_dl_9} employ a diffusion model and knowledge distillation to produce higher-quality depth images, surpassing those generated by the teacher network. Nevertheless, the aforementioned studies primarily focus on network modifications or surface-level issues, lacking a thorough investigation into the core process of features decoding for depth estimation.\par 

Unlike previous studies that treat depth estimation as a mere modality transformation, we posit the existence of a space that represents physical entities in endoscopic surgery. By validating this space and extracting its intrinsic features, depth interpretation can be achieved with greater accuracy in endoscopic image depth estimation.\par

\section{Methodology} 

\begin{figure}[t]
	\centering    
	\includegraphics[width=1.0\linewidth]{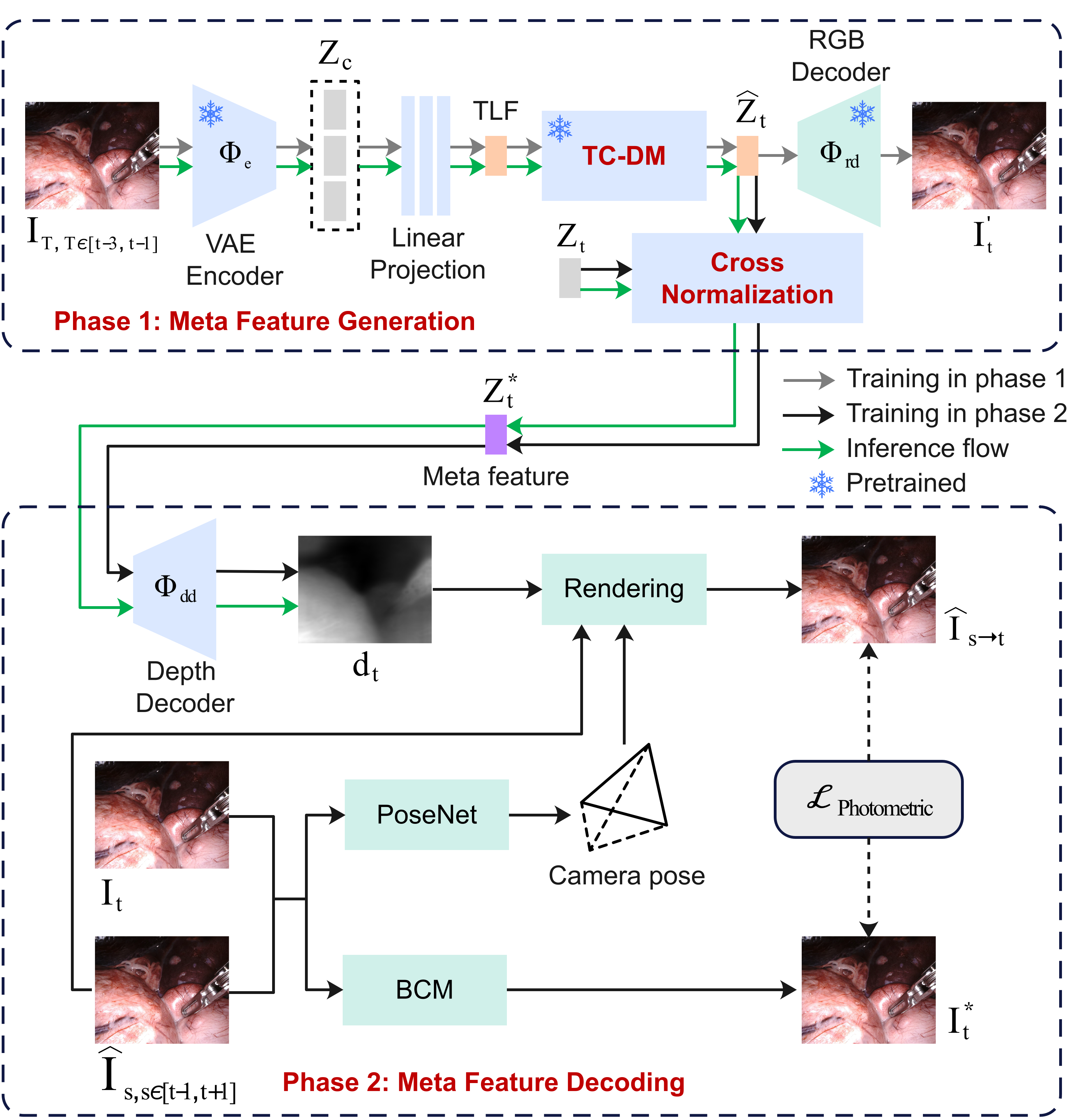}
	\caption{The structure of the proposed framework (MetaFE-DE), which consists of the two phases, i.e., meta feature generation and decoding.}      
	\label{fig:method_framework}  
\end{figure}  

The proposed MetaFE-DE, as shown in Fig. \ref{fig:method_framework}, comprises two phases, i.e., meta feature generation and meta feature decoding. In the first phase, meta-features $Z^{*}_{t}$ are generated by leveraging the diffusion process, pixel-wise self-supervised pre-training, and features alignment across spatial and temporal spaces. In the second stage, $Z^{*}_{t}$ is decomposed into the depth image using the self-supervised learning framework based on a classical brightness-calibration monocular depth estimation approach \cite{rw_dl_7}. 

\subsection{Phase 1: Meta Feature Generation}

Entities in the physical-world inherently exhibit both spatial and temporal features. While vanilla diffusion models are inherently suited to capture spatial information during training, they lack effective integration of temporal dynamics. To address this limitation, this study employs sequential images to encode temporal information, which is referred as temporal latent feature ($\mathrm{TLF}$) in this paper (Section \ref{TLF}), thereby enhancing the generation of meta-features. The $\mathrm{TLF}$ is then taken as temporal cues for the temporal conditioned diffusion module (TC-DM) (Section \ref{TC-DM}), which generates the latent diffusion features $\hat{Z}_{t}$. To effectively align and integrate the spatiotemporal features of the current frame within the latent space, we employ cross-normalization (Section \ref{CN}) to ensure consistency in their distributions. Ultimately, aligned features $Z^{*}_{t}$ is defined as the meta features in this paper.

\subsubsection{Temporal Latent Feature Learning}
\label{TLF}
As shown in \eqref{eq1}, each image in the sequence ${I_{\mathrm{T}, T \in [t-3, t-1]}}$, where $I_{t}\in \mathbb{R}^{H \times W}$, is independently processed by the VAE encoder $\Phi_{e}$, resulting in latent features ${Z_{\mathrm{T}, T \in [t-3, t-1]}}$, where $Z_{t}\in \mathbb{R}^{m \times n}$. ${Z_{\mathrm{T}}}$ are then concatenated into $Z_{c} \in \mathbb{R}^{ 3 \times m \times n}$. To align these concatenated features with $Z_{t}$, a linear projection $\mathcal{F}$ is applied to transform them into $\mathrm{TLF} \in \mathbb{R}^{m \times n}$. Since each element in ${Z_{\mathrm{T}}}$ resides in the latent space, using a linear mapping helps preserve the latent spatial distribution. 
\begin{equation} 
	\mathrm{TLF} = \mathcal{F}(\mathrm{Concat}(\Phi_{e}(I_{\mathrm{T}}))), T \in [t-3, t-1].
	\label{eq1}
\end{equation} 

\subsubsection{Temporal Conditioned-Diffusion Model}
\label{TC-DM}
By infusing additional information into each denoising step, the conditioned latent diffusion  model \cite{con_1,con_2,con_3} forces intrinsic features closely align with desired output features. The objective function of the proposed TC-DM defined as

\begin{equation}  
	L_{\mathrm{CDM}}:=\mathbb{E}_{\mathcal{E}(x),y,\epsilon\sim\mathcal{N}(0,1),t}\bigg[\|\epsilon\!-\!\epsilon_\theta(z_t,t,\mathrm{TLF})\|_2^2\bigg],  
	\label{eq2}
\end{equation}
where $z_t \in \mathbf{Z}$ represents disturbed features in latent space, $\epsilon$ and $\epsilon_\theta(z_t,t,\mathrm{TLF})$ denote the added noise and predicted noise, respectively. Using the pretrained TC-DM, we can obtain the latent diffusion features $\hat{Z}_{t}$ with additional temporal information.

\subsubsection{Cross Normalization}
\label{CN}
Considering $\hat{Z}_{t}$ is derived from the temporal information of the three preceding frames, it lacks explicit spatial information from the current frame. To further encompass the spatial information from the current frame, we leverage ${Z}_{t}$ for compensation. As $\hat{Z}_{t}$ and ${Z}_{t}$ are derived by different feature extraction schemes, namely diffusion and convolution, we need to reconcile them within a unified representational space. In this work, we employ the cross normalization, which harmonizates the distribution with no parameterization \cite{cross_norm}, to align the distributional features.  The mean and variance of $\hat{Z}_{t}$ are defined as \eqref{eq3} and \eqref{eq4}, respectively. 

\begin{equation}
	\mu_{\hat{Z}_{t}}=\frac{1}{n}\sum_{i=1}^{n}\hat{Z}_{t,i},
	\label{eq3}
\end{equation}   
 
\begin{equation}
	\sigma_{\hat{Z}_{t}}^{2}=\frac{1}{n}\sum_{i=1}^{n}(\hat{Z}_{t,i}-\mu_{\hat{Z}_{t}})^{2},
	\label{eq4}	
\end{equation}   

\begin{equation}
	Z^{*}_{t}=\frac{Z_{t}-\mu_{\hat{Z}_{t}}}{\sqrt{\sigma_{\hat{Z}_{t}}^2+\epsilon}}*\gamma + \hat{Z}_{t},
	\label{eq5}  
\end{equation}
where $n$ denotes the number of samples in a batch, $\epsilon$ represents a minor constant introduced to ensure numerical stability, and $\gamma$ serves as a scaling parameter, enabling the model to adjust the normalized values effectively. \par

\subsection{Phase 2: Meta Feature Decoding}
We posit that the meta feature $Z^{*}_{t}$ encodes the better representations that can be adapted to any modality with appropriate guidance. This study focuses on its capacity to generate depth images. Given the absence of ground truth, we employ a classic monocular depth estimation framework for image generation (Section \ref{depth_estimation}). To address illumination variations in endoscopic scenarios, we integrate the illumination correction module (Section \ref{BCM}) proposed by \cite{bright} to minimize training noise.

\subsubsection{Monocular Depth Estimation}
\label{depth_estimation}
The self-supervised scheme leverages the calculated depth and pose data as intermediaries and utilizes them to warp adjacent views into the target view to provide supervisory signals\cite{endo_depth_survey,dl_1,dl_4}. Given target frame $I_{t}\left(\mathbf{p}\right)$ and source frame $I_{s}\left(\mathbf{p}\right)$, the warping operation is defined as 
\begin{equation}
	\left.h\left(\mathbf{p}_{s\to t}\right)=\left[\mathbf{K}\left|\mathbf{0}\right.\right]\mathbf{M}_{t\to s}\left[\begin{array}{c}\mathbf{D}_{t}\mathbf{K}^{-1}h\left(\mathbf{p}_{t}\right)\\1\end{array}\right.\right],
	\label{eq6} 
\end{equation}  
where $h\left(\mathbf{p}_{s\to t}\right)$ and $h\left(\mathbf{p}_{t}\right)$ denote the homogeneous pixel coordinates in the source view $s$ and target view $t$, respectively. Here, $\mathbf{K}$ represents the camera intrinsic matrix, $\mathbf{M}_{t\to s}$ describes the ego-motion transformation from $t$ to $s$, and $\mathbf{D}_{t}(p)$ denotes the depth map, which is predicted by a depth decoder (Section \ref{sec:depth_decoder}), at pixel $p$ in the target frame $I_{t}\left(\mathbf{p}\right)$, the synthetic frame $\hat{I}_{s\to t}\left(\mathbf{p}\right)$ is generated through a differentiable inverse warping operation, implemented as a spatial transformer\cite{spatial}. 

The photometric loss $\mathcal{L}$ of such self-supervised scheme aims to force the synthesized frame $\hat{I}_{s\to t}\left(\mathbf{p}\right)$ to be the same as the original target frame $I_{t}\left(\mathbf{p}\right)$, thereby optimizing the deep decoder $\Phi_{dd}$ (Fig. \ref{fig:method_framework}) and the pose estimation network. $\mathcal{L}$ is defined as
\begin{equation}   
	\mathcal{L}\left(I_t,\hat{I}_{s\to t}\right)=\alpha\frac{1-\mathrm{SSIM}\left(I_t,\hat{I}_{s\to t}\right)}{2}+(1-\alpha)\left|I_t-\hat{I}_{s\to t}\right|,
\end{equation}
which means the training pipeline is fully supervised by the discrepancy in appearance between  $\hat{I}_{s\to t}\left(\mathbf{p}\right)$ and $I_{t}\left(\mathbf{p}\right)$. Additional constrains (e.g., smoothness) are consistent with \cite{rw_dl_8}. \par
\subsubsection{Brightness Calibration Module}
\label{BCM}
Considering the invariant illumination caused by endoscopy moving, AF-SFM framework \cite{bright} is proposed to alleviate the inconsistency of target frame $I_{t}\left(\mathbf{p}\right)$ and source frame $I_{s}\left(\mathbf{p}\right)$. Specifically, optical flow is introduced as prior-knowledge to learn the appearance residual $\mathbf{C}_{\delta}(\mathbf{p})$, then the primarily training objective is to minimize $\mathcal{L}\left(I_t,\hat{I}_{s\to t} + \mathbf{C}_{\delta}(\mathbf{p}) \right)$. In this study, we extend this self-supervised learning framework with brightness calibration to enable the decoding of meta features $Z^{*}_{t}$ into depth image. \par 

\subsubsection{Depth Decoder}
\label{sec:depth_decoder}
In this study, we employ a network with the same architecture used in the VAE decoder \cite{ldm} as the depth decoder $\Phi_{dd}$, which consists of three scales ($64 \times 80$, $128 \times 160$, and $256 \times 320$) of convolution. During training, three weight initialization methods (Section \ref{ablation}) are applied to examine the transformation of feature embedding space under different conditions and evaluate the performance of our method. For simplicity, we refer layers with the scale of $64 \times 80$ as deeper layers ($0\sim6$), the scale of $128 \times 160$ as middle layers ($7\sim11$), and th scale of $256 \times 320$ as shallow layers ($12\sim14$) (more details of VAE decoder are described in Appendix \ref{vae-decoder}).

\section{Experiments and Results}

\subsection{Datasets}
\begin{itemize}
	\item The SCARED dataset\cite{SCARED} consists of $35$ endoscopic video sequences from porcine cadavers, with ground-truth annotations for point clouds and ego-motion.
	\item The EndoSLAM dataset\cite{endo_slam} includes ex vivo porcine gastrointestinal tract organs with ground-truth depth information for endoscopic image depth estimation.
	\item The Hamlyn dataset\footnote{\url{https://hamlyn.doc.ic.ac.uk/vision/}} consists of phantom heart model videos with point cloud ground truth, as well as in vivo endoscopic videos from various surgical procedures.  
\end{itemize}   

\begin{table*}[ht] % 包装表格的环境
	\centering % 使表格居中 
	\renewcommand{\arraystretch}{1.5} % 增加每一行之间的间距
	\caption{Performance comparison for four depth estimation methods on SCARED dataset (the winner is in bold) } % 表格标题
	\label{scared}
	\footnotesize
	\begin{tabular}{c@{\hspace{6pt}}c@{\hspace{6pt}}c@{\hspace{8pt}}c@{\hspace{6pt}}c@{\hspace{8pt}}c@{\hspace{6pt}}c@{\hspace{8pt}}c@{\hspace{6pt}}c@{\hspace{8pt}}c@{\hspace{6pt}}cc@{\hspace{8pt}}c}
		\toprule % 顶部加粗横线
		\textbf{Methods} & \cellcolor{orange!20}Abs Rel $\downarrow$ & 95\%CIs & \cellcolor{orange!20}Sq Rel $\downarrow$ & 95\%CIs & \cellcolor{orange!20}RMSE $\downarrow$ & 95\%CIs & \cellcolor{orange!20}RMSE log $\downarrow$ & 95\%CIs & \cellcolor{orange!20}$\delta \uparrow$ & 95\%CIs \\
		\midrule % 第二行加粗横线
		\midrule % 第三行加粗横线
		\textbf{MonoDepth2} & 0.073 &[0.071, 0.075]& 0.626 &[0.610, 0.642]& 5.987 &[5.850, 6.124]& 0.099 &[0.097, 0.103]& 0.950 & [0.945, 0.955] \\
		\textbf{LiteMono} & 0.061 &[0.059, 0.062]& 0.472 &[0.458, 0.486]& 5.127 &[5.015, 5.239]	& 0.085 &[0.083, 0.087]& 0.967 & [0.963, 0.972]  \\
		\textbf{MonoDiffusion} & 0.060 &[0.058, 0.062]& 0.458 &[0.444, 0.472]& 5.116 &[4.996, 5.226]& 0.083 &[0.081, 0.085]& 0.969 & [0.965, 0.971] \\  
		\midrule
		\textbf{MetaFE-DE (Ours)} & \textbf{0.056} &[0.054, 0.057]& \textbf{0.423} &[0.411, 0.435]& \textbf{5.015} &[4.905, 5.125]	& \textbf{0.080} &[0.079, 0.082]	& \textbf{0.972} & [0.968, 0.976] \\
		\bottomrule % 最底部加粗横线  
	\end{tabular}
\end{table*}

\begin{table*}[ht] % 包装表格的环境
	\centering % 使表格居中
	\renewcommand{\arraystretch}{1.5} % 增加每一行之间的间距
	\caption{Performance comparison for four depth estimation methods on EndoSlam dataset (the winner is in bold) } % 表格标题
	\label{endoslam}
	\footnotesize
	\begin{tabular}{c@{\hspace{6pt}}c@{\hspace{6pt}}c@{\hspace{8pt}}c@{\hspace{6pt}}c@{\hspace{8pt}}c@{\hspace{6pt}}c@{\hspace{8pt}}c@{\hspace{6pt}}c@{\hspace{8pt}}c@{\hspace{6pt}}cc@{\hspace{8pt}}c}
		\toprule % 顶部加粗横线
		\textbf{Methods} & \cellcolor{orange!20}Abs Rel $\downarrow$ & 95\%CIs & \cellcolor{orange!20}Sq Rel $\downarrow$ & 95\%CIs & \cellcolor{orange!20}RMSE $\downarrow$ & 95\%CIs & \cellcolor{orange!20}RMSE log $\downarrow$ & 95\%CIs & \cellcolor{orange!20}$\delta \uparrow$ & 95\%CIs \\
		\midrule % 第二行加粗横线  
		\midrule % 第三行加粗横线
		\textbf{MonoDepth2} & 0.075 &[0.073,0.076]& 0.764 &[0.749,0.779]& 7.046 &[6.902,7.190]& 0.104 &[0.103,0.106]& 0.938 & [0.933,0.941] \\
		\textbf{LiteMono} & 0.072 &[0.070, 0.074]& 0.653 &[0.639, 0.667]& 6.131 &[6.002, 6.260]& 0.100 &[0.098, 0.102]& 0.948 & [0.943, 0.952] \\
		\textbf{MonoDiffusion} & 0.071 &[0.069, 0.073]& 0.631 &[0.617, 0.645]& 6.011 &[5.883, 6.139]& 0.097 &[0.095, 0.100]& 0.951 & [0.947, 0.955]  \\
		\midrule 
		\textbf{MetaFE-DE (Ours)} & \textbf{0.068} &[0.066, 0.070]& \textbf{0.614} &[0.600, 0.628]& \textbf{5.901} &[5.773, 6.029]& \textbf{0.096} &[0.094, 0.098]	& \textbf{0.956} & [0.952, 0.960]
		\\
		\bottomrule % 最底部加粗横线  
	\end{tabular}
\end{table*}
\begin{table*}[ht] % 包装表格的环境
	\centering % 使表格居中
	\renewcommand{\arraystretch}{1.5} % 增加每一行之间的间距
	\caption{Performance comparison for four depth estimation methods on Hamlyn dataset (the winner is in bold)} % 表格标题
	\label{hamlyn}
	\footnotesize
	\begin{tabular}{c@{\hspace{4pt}}c@{\hspace{4pt}}c@{\hspace{8pt}}c@{\hspace{6pt}}c@{\hspace{8pt}}c@{\hspace{4pt}}c@{\hspace{8pt}}c@{\hspace{4pt}}c@{\hspace{8pt}}c@{\hspace{4pt}}cc@{\hspace{8pt}}c} 
		\toprule % 顶部加粗横线
		\textbf{Methods} & \cellcolor{orange!20}Abs Rel $\downarrow$ & 95\%CIs & \cellcolor{orange!20}Sq Rel $\downarrow$ & 95\%CIs & \cellcolor{orange!20}RMSE $\downarrow$ & 95\%CIs & \cellcolor{orange!20}RMSE log $\downarrow$ & 95\%CIs & \cellcolor{orange!20}$\delta \uparrow$ & 95\%CIs \\
		\midrule % 第二行加粗横线  
		\midrule % 第三行加粗横线
		\textbf{MonoDepth2} & 0.092 &[0.090, 0.094]& 1.755 &[1.720, 1.786]& 13.179 &[12.950, 13.410]& 0.167 &[0.165, 0.170]& 0.881 & [0.878, 0.884]\\
		\textbf{LiteMono} & 0.089 &[0.087, 0.091]& 1.701 &[1.670, 1.732]& 13.017 &[12.780, 13.254]& 0.163 &[0.161, 0.165]& 0.885 &  [0.882, 0.887]\\
		\textbf{MonoDiffusion} & 0.089 &[0.087, 0.091]& 1.694 &[1.662, 1.726]& 12.985 &[12.750, 13.220]& 0.163 &[0.161, 0.165]& 0.886 & [0.883, 0.889] \\  
		\midrule 
		\textbf{MetaFE-DE (Ours)} & \textbf{0.071} &[0.069, 0.073]& \textbf{1.065} &[1.040, 1.090]& \textbf{10.503} &[10.320, 10.686]& \textbf{0.124} &[0.122, 0.126]& \textbf{0.946} & [0.943, 0.949]
		\\
		\bottomrule % 最底部加粗横线  
	\end{tabular}
\end{table*}
\begin{table*}[ht]
	\centering
	\renewcommand{\arraystretch}{1.5} % 增加每一行之间的间距
	\caption{Experimental results for the ablation study on SCARED dataset.} % 表格标题
	\label{ablation-results}
	\footnotesize
	\begin{tabular}{c@{\hspace{6pt}}c@{\hspace{12pt}}c@{\hspace{8pt}}c@{\hspace{6pt}}c@{\hspace{8pt}}c@{\hspace{6pt}}c@{\hspace{8pt}}c@{\hspace{6pt}}c@{\hspace{8pt}}c@{\hspace{6pt}}cc@{\hspace{8pt}}c}
		\toprule
		\textbf{WP} & \textbf{CN} & \cellcolor{orange!20}Abs Rel $\downarrow$ & 95\%CIs & \cellcolor{orange!20}Sq Rel $\downarrow$ & 95\%CIs & \cellcolor{orange!20}RMSE $\downarrow$ & 95\%CIs & \cellcolor{orange!20}RMSE log $\downarrow$ & 95\%CIs & \cellcolor{orange!20}$\delta \uparrow$ & 95\%CIs \\
		\midrule 
		\midrule
		\textbf{\checkmark} & \textbf{\checkmark} & \textbf{0.056} &[0.054, 0.057]& \textbf{0.423} &[0.411, 0.435]& 5.015 &[4.905, 5.125]	& \textbf{0.080} &[0.079, 0.082]  & \textbf{0.972} & [0.968, 0.976] \\
		\textbf{$\times$} & \textbf{\checkmark} & 0.057 &[0.056, 0.058]& 0.464 &[0.452, 0.476]& \textbf{4.979} &[4.880, 5.078]& 0.081 &[0.079, 0.082]& 0.970 & [0.968, 0.972] \\
		\textbf{\checkmark}  & \textbf{$\times$}& 0.060 &[0.059, 0.062]& 0.470 &[0.458, 0.482]& 5.120 &[5.020, 5.220]& 0.085 &[0.083, 0.087]& 0.966 & [0.963, 0.970] \\  
		 \textbf{$\times$} &\textbf{$\times$} & 0.063 &[0.061, 0.064]& 0.571 &[0.558, 0.584]& 8.983 &[8.800, 9.166]& 0.089 &[0.087, 0.092]& 0.963 & [0.961, 0.967] \\
		\bottomrule 
	\end{tabular}
	\vspace{0.5cm}  % 用于添加间距
	\textbf{Note:} WP refers to ``with the pretrained weights of RGB decoder" and CN refers to ``cross normalization".
\end{table*}
\subsection{Implementation Details}
Our framework is trained using the PyTorch \cite{pytorch} and trained on a server with four NVIDIA GeForce RTX 4090 GPUs ($24$ GB). The input resolution for all subnetworks is set to $320 \times 256$ pixels. The training process consists of two stage: the first stage follows the training process of LDM and is divided into two sub-stages. In the first sub-stage, we train the VAE for around $30$ epochs. In the second sub-stage, we train the first phase for $12$ epochs to obtain a stable $\hat{Z}$ representation. The second phase adheres to the training process of AF-Net and is also divided into two sub-stages. In the first sub-stage, we train the OF-Net for approximately $20$ epochs. In the second sub-stage, we train the depth decoder and pose-net networks for $18$ epochs. By obtaining the intrinsic attributes of the physical entity itself, we can accelerate the convergence of the depth estimation module. The metrics \cite{rw_dl_1} for assessing accuracy in depth evaluation are Abs Rel, Sql Rel, RMSE, RMSE log, and $\delta$. The SCARED dataset is split into 18670, 1000, and 300 frames for training, validation, and test sets, respectively. For EndoSlam dataset, we use the synthetic colon dataset, and split it into 18750, 1000, 300 for training, validation, and test sets, respectively. The definition of metrics,the value of hyperparameters and more evaluation details are described in Appendix \ref{metrics}.

\subsection{Performance Evaluation}
\subsubsection{Comparison Study}
We assess the depth estimation accuracy of our framework by comparing it with three related self-supervised methods, including LiteMono\cite{rw_dl_8}, MonoDiffusion\cite{rw_dl_9}, and MonoDepth2\cite{rw_dl_7}. For the results of evaluation metrics, the confidence intervals (CIs) are calculated, and the paired t-test is performed for the statistical significance validation on the improvements of performance. These methods are reproduced with the open source code. Tables \ref{scared}, \ref{endoslam}, \ref{hamlyn} present the experimental results of our method and the three compared methods on SCARED, Endo-Slam, and Hamlyn dataset, respectively. 
%For the results, Sq Rel is highly sensitive to substantial depth errors, Abl Rel is more sensitive to depth values in closer regions.

\begin{figure}[t]  
	\centering       
	\includegraphics[width=1.0\linewidth]{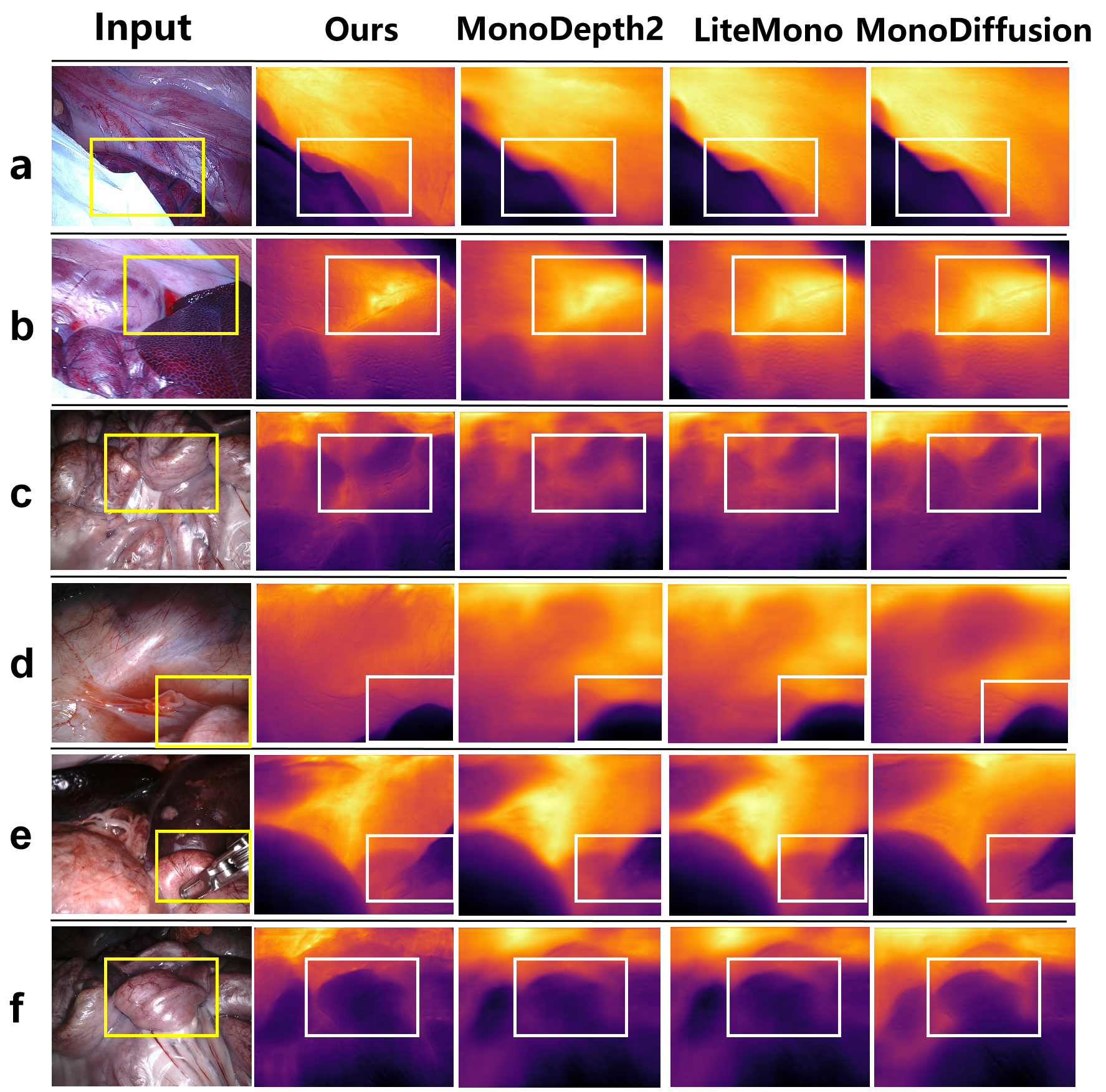}    
	\caption{By decoding the depth information from MetaFE, our method generates the depth images with more accurate details compared with three related methods.}   
	\label{fig:compare-depth}
\end{figure}    

Experimental results show that our method outperforms all compared method significantly on various endoscopic datasets ($p < 0.05$). In general, failing to account for brightness inconsistencies in endoscopic surgery scenes, Monodepth2 \cite{rw_dl_7} presents inferior performance compared to other methods. LiteMono \cite{rw_dl_8} simultaneously considers lightweighting and illumination correction, demonstrating a significantly enhanced performance compared to Monodepth2 \cite{rw_dl_7}. Owing to the denoising ability of diffusion model, MonoDiffusion \cite{rw_dl_9} facilitates the acquisition of a more accurate depth image. Nevertheless, the optimization capacity is constrained by the limitations of the pseudo ground truth. \par

Table \ref{scared} indicates that our method achieves an Abs Rel of $0.056$, significantly lower than the $0.06$ obtained by MonoDiffusion, demonstrating the superior performance achieved by our method in depth prediction for nearby areas. This is further illustrated in Fig. \ref{fig:compare-depth}, where the first, fourth, and fifth rows show that tissues and surgical instruments closer to the camera exhibit enhanced depth details and sharper object edges. Additionally, the second row reveals that even distant areas can present a clearer depth effect. Across all metrics on the SCARED dataset, our method outperforms existing approaches, as shown in Table \ref{scared}. Figure \ref{fig:compare-depth} also highlights that the depth image generated by our method shows  clearer and more detailed depth representations compared to the three competing methods. The Endo-Slam dataset faces significant image homogeneity challenges. As indicated in Table \ref{endoslam}, our approach surpasses all related methods methods, suggesting that the proposed meta features from modality generation tasks provide benefits over traditional features from modality conversion, thus alleviating the effects of sparse textures in endoscopic images on depth estimation. To further validate the generalization of our method with MetaFE, we directly use the weights trained on the SCARED dataset for depth estimation on the Hamlyn dataset. Surprisingly, as shown in Table \ref{hamlyn}, our method achieves significant improvements without fine-tuning, with Abs Rel and Sq Rel metrics of $0.071$ and $1.065$, respectively. In comparison, MonoDiffusion\cite{rw_dl_9} yields metrics of $0.089$ and $1.755$. These results verify the generalization of the learned MetaFE across different datasets and, showcasing its superior performance over SOTA on an untrained dataset.

 To further investigate the feature embedding space transformation during the decoding of meta features into RGB and depth images, we evaluate the similarity between features generated by the RGB and depth decoders using CKA (see Appendix \ref{meta-feature-decoding}). Notion that we directly calculate CKA values for network layers of the same scale, for layers with differing scales, we first use the principal component analysis (PCA) to align them,and then proceed with CKA computation.

\begin{figure}[t] 
	\centering       
	\includegraphics[width=1.0\linewidth]{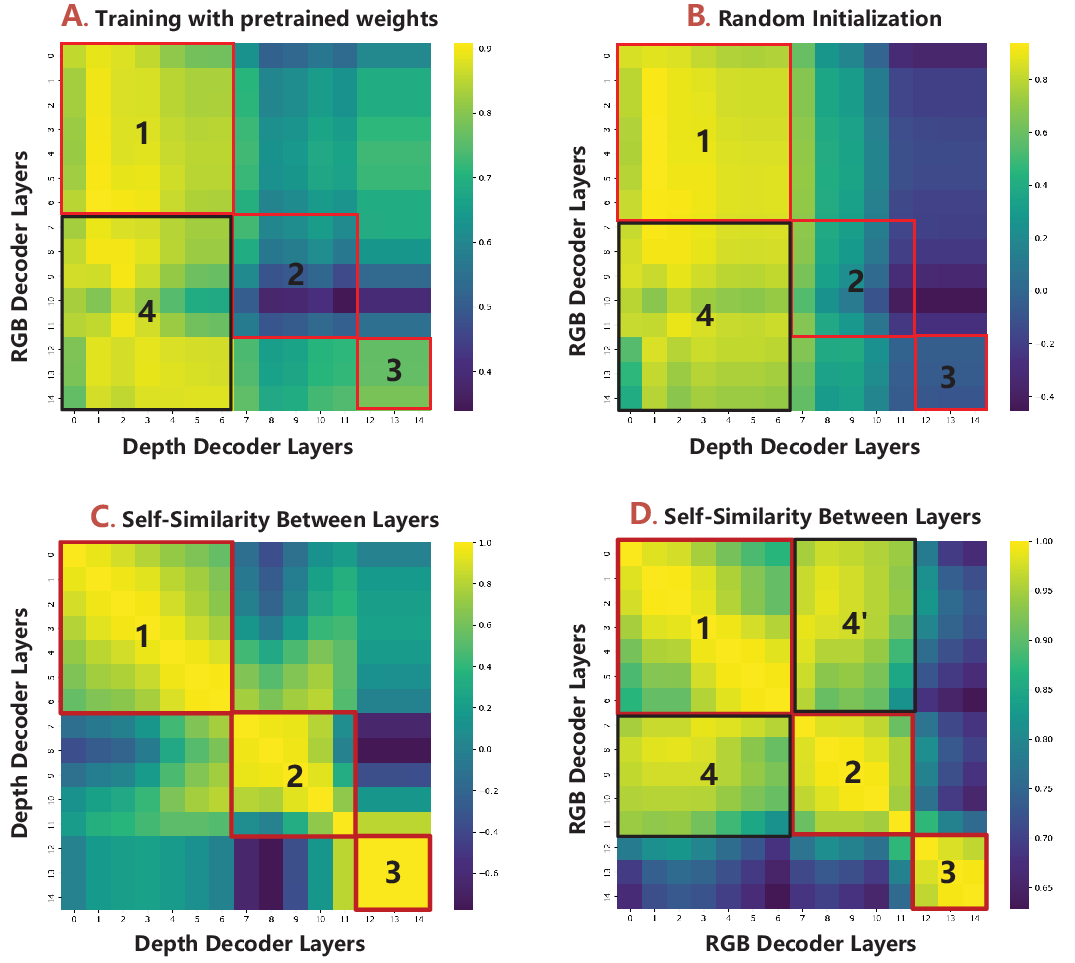}    
	\caption{Feature similarity using CKA, with axes representing network layers and cell values indicating similarity. A: Similarity at each layer between the depth and RGB decoders (depth decoder trained with RGB pre-trained weights). B: Similarity at each layer between the depth decoder (trained from scratch) and the RGB decoder. C: Intra-layer similarity within the depth decoder (trained from scratch). D: Intra-layer similarity within the RGB decoder.}     
	\label{fig:CKA}  
\end{figure}

In Fig. \ref{fig:CKA}, The horizontal and vertical axes corresponding to the features of each network layer, while the coordinate values reflect the similarity between these features. ``RGB Decoder Layers" refers to decoder layers with pre-trained weights obtained from the current RGB frame generation task(Fig. \ref{fig:method_framework}-Phase 1), ``Depth Decoder Layers" represents decoder layers with weights for depth estimation (Fig. \ref{fig:method_framework}-Phase 2). It should be noted that the weight initialization strategy differs when training the depth decoder. In Fig. \ref{fig:CKA}A, the weights of depth decoder are initialized with the pre-trained RGB decoder, while in  Fig. \ref{fig:CKA}B and \ref{fig:CKA}C, the depth decoder is initialized randomly. \par

Fig. \ref{fig:CKA}A and \ref{fig:CKA}B consistently reveal high features similarity in the deeper layers (block1), suggesting that depth and RGB information become spatially aligned. The middle layers show pronounced features distinction (block2), demonstrating their critical role in differentiating depth from RGB representations. The different CKA values in shallow layers (block3) show more reliance on the pre-trained weights. Besides, features of depth decoder layers also show a notable similarity to those in the RGB decoder (block4), not only at the same scale but also across other layers. Given our hypothesis that meta features represent the same physical entity, we conject that they share weights within an abstract space when decoded across different modalities. Block1 illustrates that during meta features decoding, the shared features distribute in deeper layers, which can be referred as an abstract feature space, from which it is subsequently decoded into RGB and depth images through different paths. In general, the similarity of features derived from the same source meta features ($Z^{*}_t$) in block 1 suggests that meta features represent intrinsic features shared by both RGB and depth images. These features can be hierarchically decomposed, layer by layer, to further reveal our hypothesis.\par     

Fig. \ref{fig:CKA}C and \ref{fig:CKA}D respectively illustrate the similarity of features between layers of the depth decoder and the RGB decoder. Fig. \ref{fig:CKA}C shows that features within depth docoder layers at the same scale exhibit clear similarity, while Fig. \ref{fig:CKA}D highlights features similarity between deeper and middle layers (block 4, 4'), indicating that during RGB image decoding, features in both layers share weights. 

\subsubsection{Ablation Study}
\label{ablation}
Given that the features in MetaFE can be decoded into RGB and depth images, it is anticipated that the decoder can work with or without pre-trained VEA decoder weights of RGB reconstruction during the depth decoder training. Therefore, we posit that weights initialization strategy is crucial for studying how meta features are decoded into depth images. Additionally, the necessity of cross normalization is also validated through the ablation study.

We designate the ablation study as follows: depth decoder initialized with or without pre-trained weights of RGB reconstruction, depth decoder initialized with fixed weights of RGB reconstruction of deeper layers ($64 \times 80$) and decoding depth from $\hat{Z}_{t}$ directly without performing cross normalization with $Z_{t}$. For simplicity, we refer ``depth decoder initialized with pre-trained weights of RGB reconstruction" as ``WP" in Table \ref{ablation-results}, and cross normalization as ``CN".  

Table \ref{ablation-results} consistently verifies the necessity of cross normalization in feature alignment. Although $\hat{Z}_{t}$ can accurately reconstruct the RGB image using input from the three preceding frames, it provides ample temporal information but lacks the spatial context of the current frame. Thus, alignment and fusion with spatial information are intrinsic to ensure information completeness. Furthermore, with cross-normalization in place, there is no significant difference in depth estimation performance, regardless of whether pre-trained weights from RGB reconstruction are used.

\section{Discussion} 

\textbf{Spatial-temporal alignment works:} In this study, we utilize cross normalization to align the temporal diffusion features $\hat{Z}_{t}$ and spatial features $Z_{t}$, but why these features can be aligned through such a method with the relately simple operations? This is attributed to our framework design, where the generation task incorporates the three preceding consecutive frames. Consequently, the generated (current) frame is essentially represented by its preceding frames through the diffusion process, which in turn allows the output to align accurately with the spatial features of the current frame.

\textbf{Meta feature is decoded into depth directly:} Practically, meta features can be reconstructed into RGB images, which in turn can be used to predict depth images. However, Fig. \ref{fig:CKA} shows that meta features can be decoded into depth image directly with high-similarity of weights in deeper layers of the pre-trained RGB decoder. This suggests meta features do not need to be fully converted into RGB images, they can be decoded into depth after passing through a common space (see Fig. \ref{fig:common-space} in Appendix). Additionally, we employ the VAE decoder architecture to reconstruct both RGB and depth images, ensuring a simplicity of the design. However, our experiments reveal that RGB and depth generation share common features even during the decoding process (Fig. \ref{fig:CKA}A, \ref{fig:CKA}B, Block 1). Specifically, the depth decoder has learned features for RGB reconstruction at scales 0-6 (Fig. \ref{fig:CKA}A, \ref{fig:CKA}B, Block 4). Thus, we believe a more streamlined approach exists, for instant, employing a lightweight network to directly map meta features to depth image or even other modality. We will reserve this investigation for future work.

\textbf{The meta features are portable:} This work addresses the absence of ground truth, where the decoder learning is guided by another self-supervised learning approach. Notably, the same approach can be applied when ground truth is used to guide the decoder through supervised learning. The advantage of this is that it allows for the full utilization of the pre-trained diffusion models, with the focus shifted to the design and learning of the decoder for any required down stream tasks.

\section{Conclusion} 

Given the complexity of endoscopic surgical scenes and the challenges in depth estimation from monocular endoscopic images, we propose a novel depth estimation method that learns meta features based on a diffusion model, enabling decoding into different modalities (RGB and depth) for various dense prediction tasks. Our approach is based on the hypothesis that a unified representation exists across different modalities derived from the same physical scene. Extensive experiments on diverse endoscopic datasets demonstrate that our method achieves accurate depth estimation and outperforms the state-of-the-art method for monocular endoscopic images. Furthermore, we show that in our method, different visual tasks (such as reconstructing RGB or depth image) share common features within the abstract layers and can be processed through a single decoder path.

%-------------------------------------------------------------------------
%%%%%%%%% REFERENCES
{\small
	\bibliographystyle{unsrt}
	\bibliography{reference}
}

\newpage  
\clearpage
\appendix
\section*{Appendix}

\section{Preliminary Experiments}

\subsection{Experimental Designate}

To evaluate the impact of introducing depth information as a condition on the RGB image generation task, we design a pre-experiment: generating RGB image with/without depth information as condition. For the first strategy, as shown in Fig. \ref{fig:pre_ex}, we concatenate three previous RGB images with the depth image of the current frame, and feed them into the ResNet encoder to obtain the features that match the dimensions of the latent features.
For the second strategy, as presented in Fig. \ref{fig:pre_ex_2}, the RGB images are generated without depth information. The preliminary experiment is conducted on the SCARED dataset, with the split of the training, validation, and test sets consistent with what is mentioned in the main text. Fig. \ref{fig:compare_depth} demonstrates that using depth images as conditions yields the images with higher quality. The experimental results are summarized in Table \ref{fig:with_depth_condition}, which presents the quantitative comparison of the generation results with and without depth information. Compared with the method without depth infromation, the RGB images yeilded with depth information are generally brighter and show superior performance across FID, PSNR, and SSIM, indicating the improved image quality.
%It is worth noting that relying solely on similarity to the original image might not be sufficient for evaluating image quality. 

\begin{center}
	\includegraphics[width=1.0\linewidth]{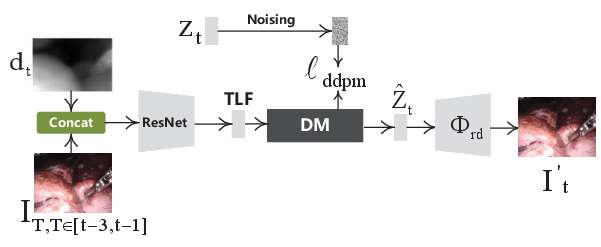}
	\captionof{figure}{The network architecture for the method using the depth image for the RGB image generation.}  % 使用 \captionof
	\label{fig:pre_ex}
\end{center}

\begin{center}
	\includegraphics[width=1.0\linewidth]{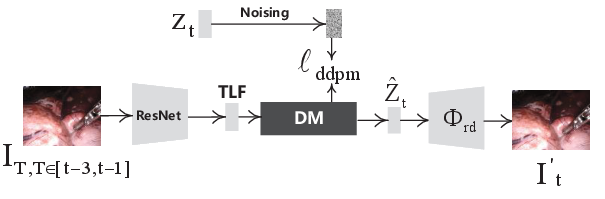}
	\captionof{figure}{The network architecture for the method without the depth image for the RGB image generation}  % 使用 \captionof
	\label{fig:pre_ex_2}
\end{center}
\subsection{Experimental Results}
\begin{center}
	\includegraphics[width=1.0\linewidth]{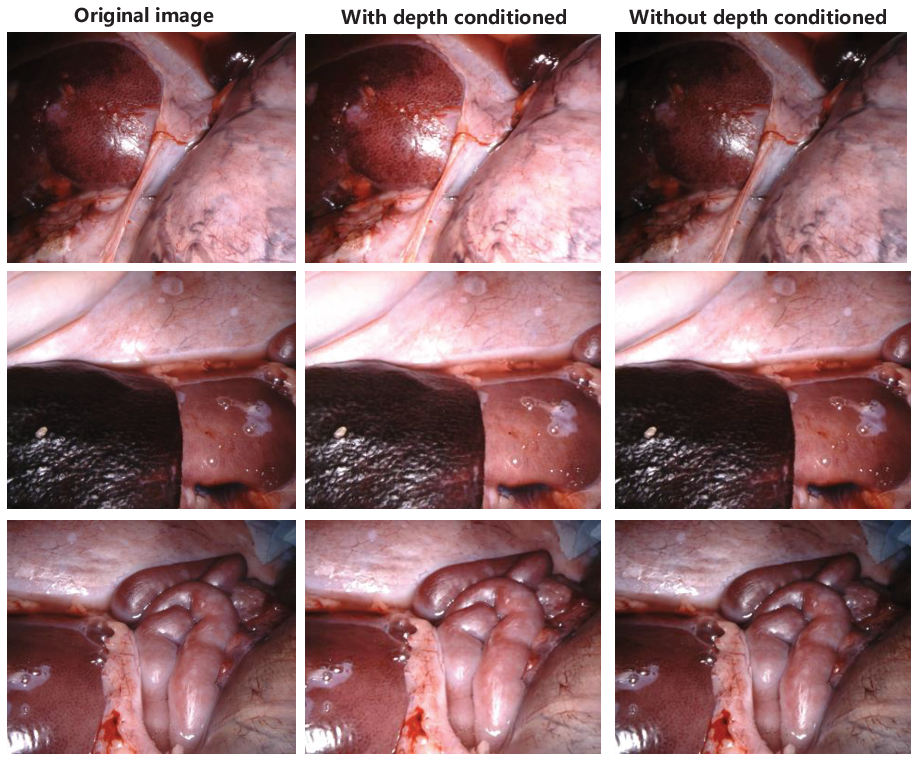}
	\captionof{figure}{The generated RGB images with and without depth information are shown in the second and third column, respectively. Notably, incorporating depth information yields the enhanced image with brighter details, surpassing its corresponding original image. Conversely, omitting depth information results in a darker image with the reduced image quality.}  % 使用 \captionof
	\label{fig:compare_depth}  
\end{center}
\begin{table}[h]
	\caption{Quantitative assessment results (mean and stand deviation) for the quality of the RGB images generated with and without depth information.}
	\label{fig:with_depth_condition} 
	\footnotesize % 或者使用 \scriptsize
	\begin{tabular*}{0.5\textwidth}{@{\extracolsep{\fill}} cccc}
		\specialrule{1.5pt}{0pt}{0pt} % 自定义粗横线，1.5pt
		& \textbf{FID} $\downarrow$ & \textbf{PSNR} $\uparrow$ & \textbf{SSIM} $\uparrow$ \\ % Bold header
		\midrule % Thin line below the header  
		w/ depth & 12.1915 $\pm$ 0.523 & 16.38991 $\pm$ 0.271 & 0.45822 $\pm$ 0.012 \\ 
		w/o depth & 15.74215 $\pm$ 0.619 & 15.9265 $\pm$ 0.305 & 0.4574 $\pm$ 0.014 \\ 
		\bottomrule % Thick bottom line
	\end{tabular*}
\end{table}

\section{Related Works}
\subsection{Diffusion Model and Representation Learning}
\label{rw1}
Mukhopadhyay et al.\cite{dif_rep_cla_1} adjust the diffusing step and the size of feature pooling to learn better feature representation for the image classification. Deja et al.\cite{dif_rep_cla_2} jointly train the generation and classification tasks by sharing weights within the diffusion process. Tian et al.\cite{dif_rep_seg_1} refer the latent mask diffusion model as a representation generator, and the segmentation network is designed to decode the representation into semantic segmentations. Zhao et al.\cite{dif_rep_multi_task} focus on training multiple down-stream tasks, such as the semantic segmentation and the depth estimation, by incorporating the text information. Li et al.\cite{dif_rep_seg_2} utilize a distillation learning framework to transfer the knowledge learned in the diffusion model to the original semantic segmentation model. Similarly, Yang et al.\cite{dif_rep} distill the features of the diffusion model into a pretrained traditional detection framework. Tian et al.\cite{dif_rep_rcg_1} utilize the representations learned from the diffusion model to enhance the recognition task performance. The aim of the aforementioned studies is to optimize the use of representations learned by the diffusion model to enhance the downstream task performance, and their experimental results consistently demonstrate the effectiveness of this strategy. 

\subsection{Modality Alignment}
\label{model-alignment}
Richens et al.\cite{real_world_3} address the problem of alignment in the perspective of the causal inference, demonstrating that the agent must have learned a causal model, thus can generalize effectively to new domains. Sharma et al.\cite{real_world_4} argue that large language models (LLMs) can generate effective visual representations from images, which are created through querying the code in LLM. All these studies demonstrate the potential for the modality alignment, thus providing a reasonable explanation for why features extracted from image generation tasks can be used to train downstream tasks.

\section{Methodology}
\subsection{VAE Decoder}
\label{vae-decoder}
\begin{table}[h]
	\centering
	\renewcommand{\arraystretch}{1.5} % 增加每一行之间的间距
	\caption{Details for the architecture of VAE decoder} 
	\footnotesize % 或者使用 \scriptsize
	\begin{tabular}{c c c c}
		\toprule
		& Deeper      & Middle        & Shallow       \\ 
		\midrule
		Feature scale & $64\times80$ & $128\times160$ & $256\times320$ \\ 
		\midrule
		Kernel size   & 3$\times$3   & 3$\times$3     & 3$\times$3     \\ 
		\midrule
		Layers        & 0 $\sim$ 6   & 7 $\sim$ 11    & 12 $\sim$ 14   \\ 
		\bottomrule % 底部加粗横线
	\end{tabular}
\end{table}

\subsection{Regularization Terms in Depth Decoding}

Consistent with \cite{bright}, in our self-supervised depth decoding framework, we include the following loss terms in addition to the photometric loss mentioned in the main text:

\begin{equation}
	\mathcal{L}_{\text{rs}}=\sum_{\mathbf{p}}|\nabla\mathbf{C}_{\delta}\left(\mathbf{p}\right)|*e^{-\nabla\left|I^{\prime}(\mathbf{p})-I^{s\to t}(\mathbf{p})\right|},
	\label{eq:Residual-based smoothness loss}
\end{equation}

\begin{equation}
	\mathcal{L}_{\text{ax}}=\sum_{\mathbf{p}}\mathbf{V}\left(\mathbf{p}\right)*\Phi\left(I^{s\to t}\left(\mathbf{p}\right),I^{t}\left(\mathbf{p}\right)+\mathbf{C}_{\delta}\left(\mathbf{p}\right)\right),
	\label{eq:Auxiliary loss}
\end{equation}

\begin{equation}
	\mathcal{L}_{\text{es}}=\sum_{p}|\nabla\mathbf{D}\left(\mathbf{p}\right)|*e^{-\nabla\left|I^{\prime}(\mathbf{p})\right|},
	\label{eq:Edge-aware smoothness loss}
\end{equation}
where \eqref{eq:Residual-based smoothness loss} constrains the smoothness of the appearance flow field, \eqref{eq:Auxiliary loss} is defined to provide an auxiliary supervisory signal for the AFNet\cite{bright}, \eqref{eq:Edge-aware smoothness loss} is utilized to constrain the property of the depth image. Since our focus is not on brightness calibration, therefore, this paper does not elaborate on the specific definitions of each loss function, for the detailed information, please refer to\cite{bright}. In general, the final loss function of depth decoding phase is defined as:

\begin{equation}
	\mathcal{L}_{\text{all}}=\mathcal{L}+\kappa\mathcal{R},
	\label{eq:final_loss}
\end{equation}
where $\mathcal{L}$ is defined in \ref{BCM}, and $\mathcal{R}\left(\mathbf{p}\right)$ is defined as:
\begin{equation}
	\mathcal{R}=\lambda_{1}\mathcal{L}_{\text{rs}}+\lambda_{2}\mathcal{L}_{\text{ax}}+\lambda_{3}\mathcal{L}_{\text{es}}.
	\label{eq:rp}
\end{equation}

\section{Experimental Details}
\subsection{Evaluation Metrics}
\label{metrics}
The evaluation metrics defined as follows:
\begin{equation}
	\text{Abs Rel} = \frac{1}{|\mathbf{D}|} \sum\limits_{d \in \mathbf{D}} \frac{|d^* - d|}{d^*},
\end{equation}

\begin{equation}
	\text{Sq Rel} = \frac{1}{|\mathbf{D}|} \sum\limits_{d \in \mathbf{D}} \frac{|d^* - d|^2}{d^*},
\end{equation}

\begin{equation}
	\text{RMSE} = \sqrt{\frac{1}{|\mathbf{D}|} \sum\limits_{d \in \mathbf{D}} |d^* - d|^2},
\end{equation}

\begin{equation}
	\text{RMSE log} = \sqrt{\frac{1}{|\mathbf{D}|} \sum\limits_{d \in \mathbf{D}} \left( \log d^* - \log d \right)^2},
\end{equation}

\begin{equation}
	\delta = \frac{1}{|\mathbf{D}|} \left| \left\{ d \in \mathbf{D} \; \middle| \; \max\left(\frac{d^*}{d}, \frac{d}{d^*}\right) < 1.25 \right\} \right| \times 100\%,
\end{equation} 
where $d$ represents the predicted depth value, and $d^{*}$ denotes the corresponding ground truth. The symbol $\mathbf{D}$ represents the collection of predicted depth values. 
In the inference phase, we apply the median scaling \cite{dl_4} to the predicted depth maps as follows:
\begin{equation}
	\mathbf{D}_{\text{scaled}}=\begin{pmatrix}\mathbf{D}_{\text{pred}}*\begin{pmatrix}\text{Median}\begin{pmatrix}\mathbf{D}_{\text{gt}}\end{pmatrix}/\text{Median}\begin{pmatrix}\mathbf{D}_{\text{pred}}\end{pmatrix}\end{pmatrix}\end{pmatrix}.
\end{equation}
The scaled depth maps are capped at $150$ mm in the SCARED and Hamlyn dataset. A range of $150$ mm and $180$ mm can cover almost all depth values. 

\subsection{Hyperparameter Settings}
The method in this study is configured with the following hyperparameters: $k = 1$, $\lambda1 = 0.01$, $\lambda2 = 0.01$, and $\lambda3 = 0.0001$, $\gamma = 0.5$, $\epsilon = 1$, the learning rate is set to $1e-4$, and the batch size is set to $16$.
\subsection{Meta Feature Decoding}
\label{meta-feature-decoding}

\begin{figure}[ht] 
	\centering       
	\includegraphics[width=1.0\linewidth]{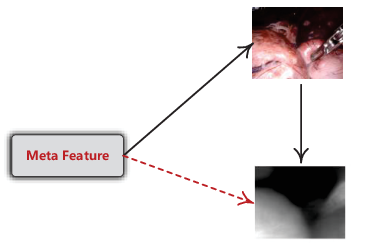}    
	\caption{The meta feature is directly decoded into the depth image, without the necessity of decoding into the RGB image in the first place.}     
	\label{fig:common-space}
\end{figure}

We provide the further explanation of our hypothesis in this study. The existence of meta features have been validated, demonstrating that these features can be decoded into either the depth or RGB image directly. We confirmed that the meta features do not need to be decoded into depth images before being transformed into RGB images, i.e., they are directly decoded into the depth images (see Fig. \ref{fig:common-space}).

\begin{figure*}[t]  
	\centering       
	\includegraphics[width=1.0\linewidth]{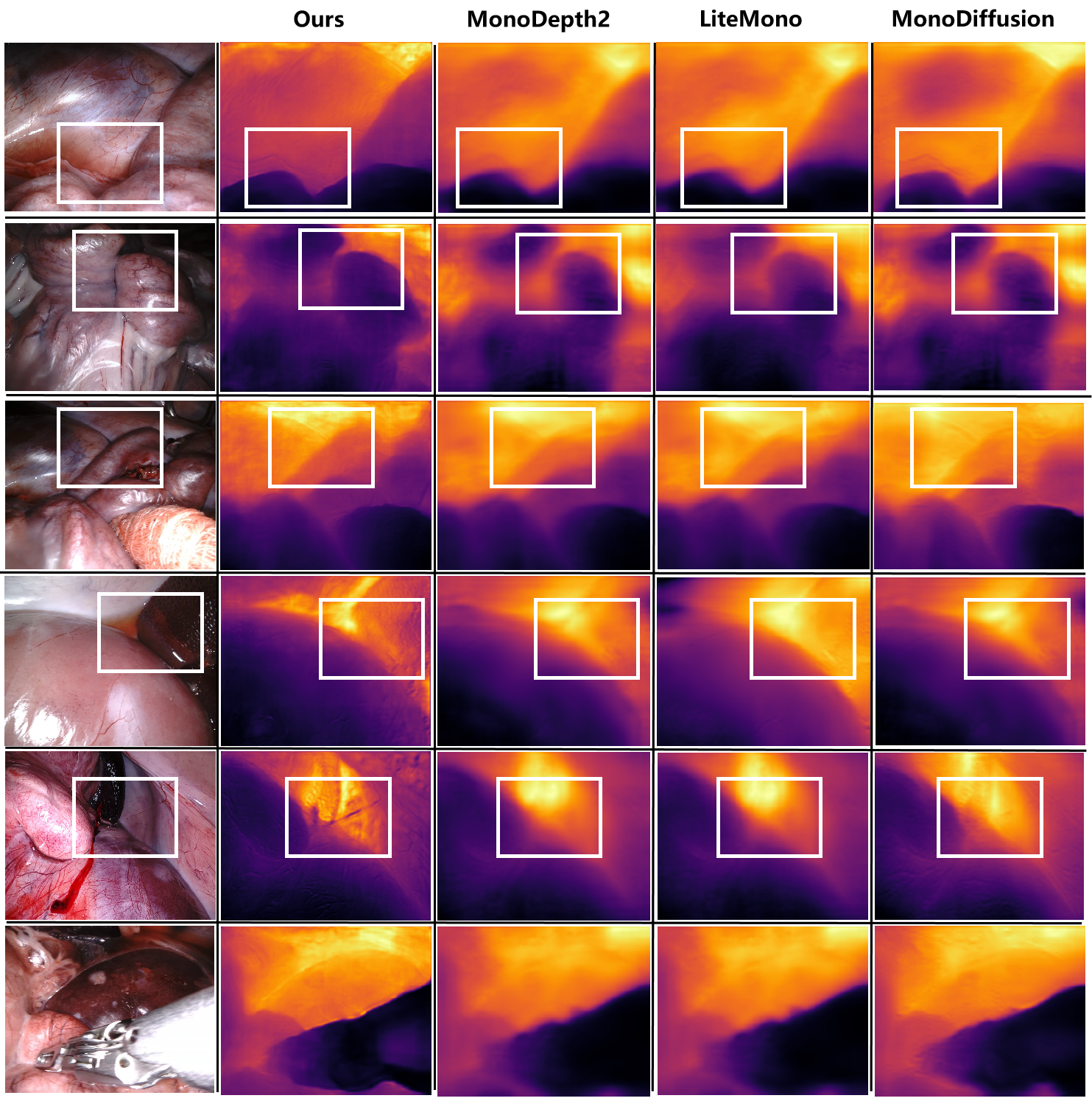}    
	\caption{More depth estimation examples on SCARED dataset.}   
	\label{fig:appen_compare-depth}
\end{figure*} 

\end{document}